\documentclass[twocolumn,showpacs,preprintnumbers,amsmath,amssymb]{revtex4}
\usepackage{graphicx,epsfig}
\usepackage{dcolumn}
\usepackage{bm}

\begin{document}

\title{Origin of the Sub-diffusive Behavior and Crossover From a Sub-diffusive to a 
Super-diffusive Dynamics Near a Biological Surface }

\author{Arnab Mukherjee\footnote[2]{email : arnab@sscu.iisc.ernet.in} and Biman 
Bagchi\footnote[1]{email : bbagchi@sscu.iisc.ernet.in}}
\affiliation{Solid State and Structural Chemistry Unit,
Indian Institute of Science,
Bangalore, India 560 012.}

\date{\today}

\begin{abstract}
 Diffusion of a tagged particle near a constraining biological surface is examined 
numerically by modeling the surface-water interaction by an effective potential. 
The effective potential is assumed to be given by an asymmetric 
double well constrained by a repulsive surface towards 
$r=0$ and unbound at large distances. The time and space dependent probability 
distribution $P(r,t)$ of the underlying Smoluchowski equation is solved by using 
Crank-Nicholson method. The mean square displacement shows a transition 
from sub-diffusive (exponent $\alpha \sim$ 0.43) to a super-diffusive 
(exponent $\alpha \sim$ 1.75) behavior with time and ultimately to a diffusive dynamics. 
The decay of self intermediate scattering function ($F_{s}(k,t)$) is non-exponential 
in general and  shows a power law behavior at the intermediate time. Such features 
have been observed in several recent computer simulation studies on dynamics of
water in protein and micellar hydration shell. The present analysis provides a simple 
microscopic explanation for the transition from the  sub-diffusivity and super-diffusivity. 
{\em The super-diffusive behavior is due to escape from the well near the surface and the 
sub-diffusive behavior is due to return of quasi-free molecules to form the bound state 
again, after the initial escape}.
\end{abstract}

\maketitle
\section{Introduction}
 The dynamics of the water molecules in the vicinity of a biological surface 
is found to be different from those in the bulk $[1-10]$. The dynamical properties of 
water around a protein surface are found to depend on the distance from the 
biomolecular surface \cite{lounnas,bizzarri1,bizzarri2,levitt}. Similar observations have 
been obtained from the computer simulations of water near micellar surface \cite{bala}. 
In particular, the mean square displacement ($MSD$) of water molecules close to biological 
surfaces is found to be sublinear with time \cite{bizzarri1,bizzarri2,cannis}. 
These results have been confirmed by neutron scattering experiments also \cite{gdsmith}. 
$MSD$ can in general be written as below,
\begin{equation}
MSD = \langle |\Delta r(t)|^{2} \rangle = c + a * t^{\alpha}
\end{equation} 
\noindent where $|\Delta r(t)|$ is the displacement of a molecule at time $t$. For 
diffusive dynamics, $\alpha = 1$. For sub-diffusive dynamics, $\alpha < 1$ whereas the 
dynamics is called super-diffusive if $\alpha > 1$. Recent computer simulation studies by 
Cannistraro {\it et al.} showed that the dynamics of water near a constrained biological 
surface is sub-diffusive for short time (below 10 ps) with the exponent $\alpha$ 
varying from 0.75 to 0.96 depending on the region of water molecules chosen at different 
distances from the protein surface \cite{rocchi,bizzarri1}. Sub-diffusivity has been 
observed in many other type of systems also, such as membrane bound protein \cite{favard}, 
polymer melts \cite{gdsmith,paul,kopf,paul1}, and the transverse motion of membrane 
point due to the bending modes \cite{granek} etc. On the other hand, enhanced or 
super-diffusion with an exponent $1.5$ has been observed in the mean square displacement 
of engulfed microsphere in a living eukaryotic cell \cite{caspi,zilman}, with a 
sub-diffusive behavior at short times \cite{salman}. The occurrence of super-diffusion has 
been interpreted as the evidence for a generalized Einstein relation, whereby 
the motion of the particles along microtubules in a dense network require displacement of 
the surrounding filaments. The time dependent diffusion coefficient is related by 
time dependent viscosity of the medium\cite{salman}. The origin of this crossover from 
sub-diffusive dynamics at short times (t$\leq$10ps) to super-diffusive dynamics at 
intermediate time (10ps$\leq$ t$\geq$ 200ps) is not yet fully understood.

 Observations for the sub-diffusive dynamics of water near the biological surface 
have often been attributed to enhanced stability of the water molecules at the surface due 
to hydrogen bonding to the biological surface \cite{saxton,bouchaud,haus,havlin}. 
Theoretical models have been developed in terms 
of bound{$\rightleftharpoons$}free equilibration \cite{nandibag}. The evidence of such 
exchange was obtained by nuclear overhausser effect experiment \cite{gotfried}. On 
the other hand, molecular dynamics simulations of water molecules at protein and micellar 
surfaces show the existence of sizable fraction of water molecules in the layer with 
residence time larger than the rest \cite{pal}. The binding energy distribution varies from 
0.5 to 9 kcal/mole. The trajectory of individual water molecules clearly shows a 
dynamic equilibrium, between a bound state and a free state \cite{cheng,gu}. To the best of 
of our knowledge, no simple explanation has been offered for the observed super-diffusive 
behavior.

 Note that no detailed analytical/numerical solution of diffusion in an underlying model 
potential has ever been carried out, even though such a study can throw light on the origin 
of the observed behavior. However, purely analytical studies in these type of systems have 
been limited by the fact that the heterogeneous surface faced by the water molecules makes a 
general analytical solution virtually impossible. 

 It is shown in recent simulations of a micelle that the bound water 
molecules are energetically stable with a potential energy contribution of 9-12 
kcal/mole while the quasi-free water molecules which surround the bound water are only 
stable by 5-6 kcal/mole \cite{pal1}. Radial distribution function shows that the first 
shell of water molecules shows a peak around 3.5 \AA \, while the second shell shows a 
peak around 6-7 \AA \cite{pal1}. The structure of $g(r)$ suggests the existence of an 
effective binding potential which could be double well in nature because the effective 
potential is related to radial distribution function by the following well-known relation
\begin{equation}
\beta V^{eff}(r) = -\ln g(r)
\end{equation}

 Recently, Garcia and Hummer reported a study of the dynamics of cytochrome c 
in an aqueous solution by molecular dynamics simulation \cite{garcia}. They observed that 
the mean square displacement of the cytochrome c exhibits a power law dependence in time 
with an exponent around 0.5 for times shorter than 100ps, and an exponent of 1.75 for longer 
times. So they clearly observed a cross-over from sub-diffusive in the short time to 
super-diffusive 
behavior in the longer time. While the short time sub-diffusion has been explained in 
terms of multi-basin dynamics, the super-diffusion is explained in terms of the 
enhancement of sampling the configurational space due to the concerted motion of atoms.
 
 In this work, we explain the origin of {\it both} the sub- and the 
super-diffusive behavior of the biological water by studying the diffusion over simple 
double well and Morse type of potentials. The double well potential 
represents the potential environment of the water close to a biomolecule. The first 
potential energy well corresponds to the bound water molecule and the second well
 corresponds to the quasi free water molecule. The effective potential energy of 
interaction between the surface and the water molecules of course becomes negligible at a 
few layers from the surface and here the water is free and behaves like as in bulk water. 
A model of free diffusion (without any potential) is studied to compare the dynamics of 
the biological water relative to the bulk water.
We demonstrate by numerical computation of the probability density $P({\bf r},t)$ on the
potential energy landscape that this model is able to catch the dynamical anomalies present 
in the biological water. The sub-diffusive behavior is shown to originate not only because 
of the binding to the surface but also due to the bound {$\rightleftharpoons$} free  
dynamic equilibration present among the  water molecules at the surface.
$MSD$ calculated from the distribution $P({\bf r},t)$ shows a cross-over from 
sub-diffusive behavior in the short time to a super-diffusive 
nature in the long time, precisely of the type observed in computer simulations 
\cite{garcia}. However, {\em we find that a pronounced sub-diffusivity is present only  
when the exchange between the two wells is facile}. Self intermediate scattering function 
$F_{s}(k,t)$ calculated from P(r,t) also shows a marked non-exponential relaxation. 
\begin{figure}[tb]
\epsfig{file=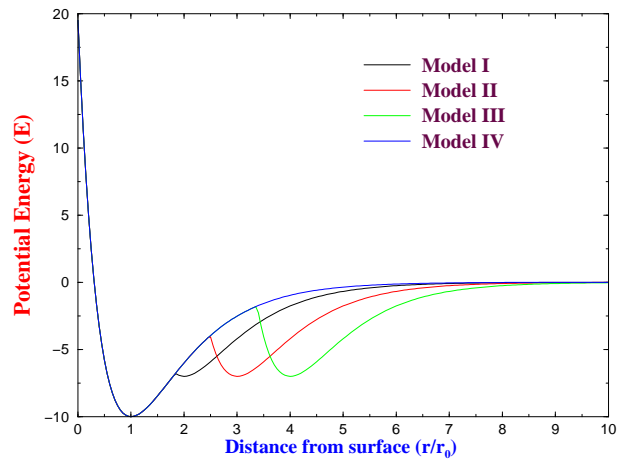,height=8cm,width=6cm,angle=-90}
\caption{Potential energy diagram for different models used in this study. Model {\bf I},
{\bf II} and {\bf III} are double well potential. Model {\bf IV} is simple Morse 
potential and Model {\bf V} is a free diffusion model without any potential.}
\end{figure}
\section{Model and Numerical Technique}
 Two Morse potentials are combined together to form different 
models of double well potentials.  The two wells are constructed in such a way so that 
the separation between the wells are $r_{0}$, 2$r_{0}$, 3$r_{0}$ in 
Model {\bf I}, Model {\bf II} and Model {\bf III}, respectively. 
In case of all the double well potentials, the depth of the first well is taken 
as -10 $k_{B}T$ and the second well is -7 $k_{B}T$. Simple Morse potential with an well 
depth of -10 $k_{B}T$ is termed as Model {\bf IV}. Simulations have also been done 
without the presence of any potential (free diffusion) which is termed as Model 
{\bf V}. Figure $1$ shows the form variation of the potentials with distances from 
a constrained surface. All the model potentials reach $0$ in the long distance. The 
double well potentials represent the stability of the first and second coordination 
shell of the water molecules surrounding the biological surface. The models have 
been constructed with different separations of double well to monitor the dynamics in 
the well and also the effect of exchange on the dynamics of water. For Model {\bf I} 
and Model {\bf II}, the two wells are close and that will facilitate the exchange 
whereas in case of Model {\bf III}, the exchange is hindered. The Morse potential 
(Model {\bf IV}), on the other hand, has no second well to perform the exchange. 
Dynamics of the probability distribution $P({\bf r},t)$ is governed by a simple 
Smoluchowski equation as given below,
\begin{equation}
{\partial \over{\partial t}} P({\bf r},t) = \nabla . D . [\nabla - \beta {\bf F}] P({\bf r},t)
\end{equation}.
\noindent where, $\beta = 1/k_{B}T$ and $F$ is the force obtained from the derivative of 
underlying potential($E$), $F=-\nabla E$. 
Without the angular dependence, the radial part of the above equation takes a simple form 
as given below, 
\begin{equation}
{{\partial P} \over {\partial t}} = D \biggl [{{\partial^{2} P} \over {\partial r^{2}}} 
+ \biggl({2\over r} - \beta {\bf F} \biggr ) {{\partial P} \over {\partial r}}  - \beta 
\biggl ({2{\bf F} \over r} + g \biggr ) P \biggr ]
\end{equation} 
\noindent where $g = \nabla.{\bf F}$. $D$ is taken to be constant. 

 The numerical solution of partial differential equation is a non-trivial task. The 
stability of the solution always depends on the equation type, variation of the 
potential and proper discretization of the equation. Here, the propagation of the 
distribution function is solved numerically by Crank-Nicholson technique \cite{crank}. 
This is an unconditionally stable method for the solution of parabolic type of equation. The 
numerical solution of a parabolic differential equation by Crank-Nicholson procedure is 
described by a simple differential equation as below,
\begin{equation}
{{\partial u} \over {\partial t}} = -{\cal H} u
\end{equation}
\noindent where ${\cal H}$ is the total Hamiltonian of the system. The formal solution 
of the above equation is,
\begin{equation}
u(t) = e^{-Ht} u(0)
\end{equation}
The above equation can be discretized as below,
\begin{equation}
u_{j}^{n+1} ={ {1-{\delta \over 2}{\cal H}} \over {1+{\delta \over 2}{\cal H}} } 
u_{j}^{n}
\end{equation}
\noindent where $n$ is the time index and $j$ is the space index. This is a half 
implicit scheme and the above equation can be solved by inverting a tridiagonal matrix. 
Although, the procedure is unconditionally stable, proper care should be taken with 
the discretization of time and distance. Force has been calculated by using cubic 
spline method where the two Morse potentials are joined to avoid the discontinuity. 
The distance $r$ is scaled by the $r_{0} = 3.5$ \AA \, where the first peak of radial 
distribution function appears for the water around a biological surface. The time $t$ 
is scaled by $\tau$ obtained from the diffusion of water ($D_{w}$) in the bulk water at 
300 K, $\tau = r_{0}^{2}/(6 D{w}) =$ 8.5 ps, where $D_{w} = 2.5 \times 
10^{-5} cm^{2}/sec$. The initial distribution of 
$P({\bf r},t)$ (at $t=0$) is taken to be a very steep function (nearly delta function) 
peaked at $r = r_{0}$. Time step $\Delta t$ is taken as $0.0001\tau$ and $\Delta r$ 
is taken as $0.05 r_{0}$. Simple explicit differentiation scheme also works in this 
case. 
\begin{figure}[tb]
\epsfig{file=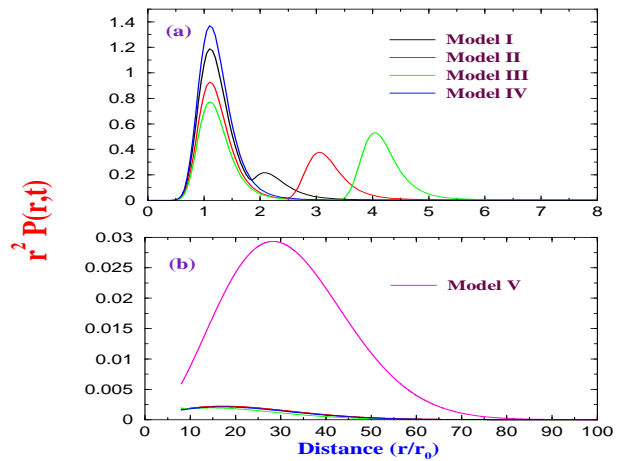,height=8cm,width=6cm,angle=-90}
\caption{(a) Probability distribution $r^{2}P(r,t)$ is plotted for short distances. All the 
models except the free diffusion shows the non-Gaussian bimodal distribution. (b) The 
distribution $r^{2}P(r,t)$ is shown for Model {\bf V} (free diffusion).}
\end{figure}
\section{Results and Discussion}
 Figure $2(a)$ shows the normalized probability distribution of 
$r^{2}\,P({\bf r},t)$ at time $t=200 \tau$ for all the model potentials. 
$P({\bf r},t)$ becomes non-Gaussian for the double and single well model potentials. 
Figure $2(b)$ shows the normalized probability distribution for the free 
diffusion model. In case of double well potential $r^{2} P(r,t)$ forms bimodal 
distribution at intermediate time. Mean square displacement is calculated from 
$r^{2}P({\bf r},t)$ using the equation below,
\begin{equation}
MSD = \langle r^{2} (t) \rangle = \int d{\bf r}\, r^{2}\, P({\bf r},t)
\end{equation}
Mean square displacement shows different extent of sub-diffusive behavior in the 
short time depending on the nature of the model potentials. The cross-over to a 
super-diffusive behavior takes place in the intermediate time and finally the dynamics 
become diffusive. This dynamical dis-symmetry is depicted in the values of 
exponent $\alpha$ obtained from the fitting of $MSD$ with Eq. $1$ 
for the different models in different time windows. Table $1$ shows the 
exponents $\alpha$ obtained for the model potentials in different time window.

\begin{center}
{\bf Table 1} 
\end{center}
\noindent The signature of the crossover from sub-diffusive to super-diffusive dynamics. 
Exponents from the fitting of $MSD$ with Eq. $1$ for different models in different 
time window. 
\begin{center}
\begin{tabular}{||c|c|c||c|c||}\hline
{\em Model} & \multicolumn{2}{c||}{\em Sub-diffusive($\alpha$)} &  
\multicolumn{2}{c||}{\em Super-diffusive($\alpha$)} \\ \cline{2-5}
          &  $0.3 \rightarrow 10\tau$ &  $0.3\rightarrow 20\tau$ & 
$20\rightarrow 200\tau$ &  $0.3\rightarrow 600\tau$\\\hline
Model I   & 0.03 & 0.96  & 1.76 & 1.65\\
Model II  & 0.58 & {\bf 0.43}  & 1.75 & 1.64\\
Model III & 0.96 & 0.97  & 1.76 & 1.81\\
Model IV  & 1.03 & 1.39  & 1.79 & 1.68\\
Model V   & 1.00 & 0.96  & 1.00 & 1.00\\\hline
\end{tabular}
\end{center}
It is clear from the values of the exponents in Table $1$ that in the short time 
(less than $10\tau$) most of the model potentials show sub-diffusive behavior. However, 
the sub-diffusivity persists for longer time (upto $20\tau$) in case of Model 
{\bf II} in which the exchange between the two wells is facilitated. All the 
models show super-diffusive dynamics in the intermediate time with exponents much 
higher than $1.0$. In the very long time (above $\approx 200\tau$), diffusive dynamics 
starts setting in. So the exponent decreases in the fit for the larger time window.
Model {\bf V} shows simple diffusion in the absence of potential with exponent always 
equal to $1.0$. Figure $3$ shows the $MSD$ against for short time of 20 $\tau$. Fitting 
of Model {\bf II} with Eq. $1$ with a very low exponent is also shown in the same 
figure. 
\begin{figure}[tb]
\epsfig{file=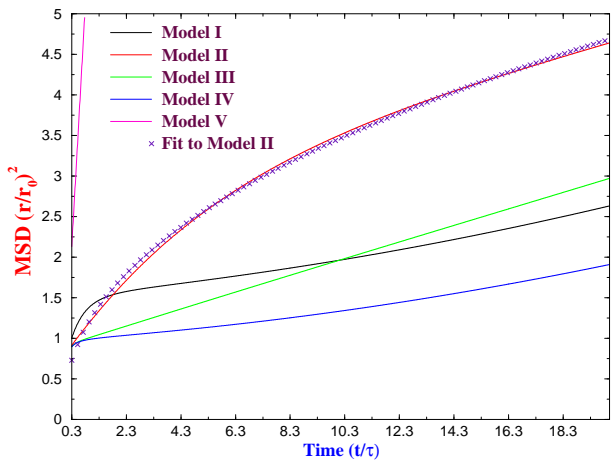,height=8cm,width=6cm,angle=-90}
\caption{Sub-diffusive behavior of $MSD$ in the short time for all the models. 
The sub-diffusive behavior is most prominent for Model {\bf II}. Free particle 
diffusion for Model {\bf V} is simple diffusive.}
\end{figure}
 Dynamics of $MSD$ changes to a super-diffusive behavior in the longer time 
(beyond $10~20 \tau$) as shown in figure $4$. All the models show a pronounced 
super-diffusive dynamics although the 
overall $MSD$ is much 
less compared to the diffusion of the free molecules as given by Model {\bf V}. 
The exponent $\alpha$ obtained from the fit of the $MSD$ are given in Table $1$. 
 
 Some nice correlations could be drawn from the change in the exponent of diffusion 
in the sub-diffusive and super-diffusive. The more pronounced sub-diffusive dynamics 
shows a less pronounced super-diffusive behavior. 
\begin{figure}[tb]
\epsfig{file=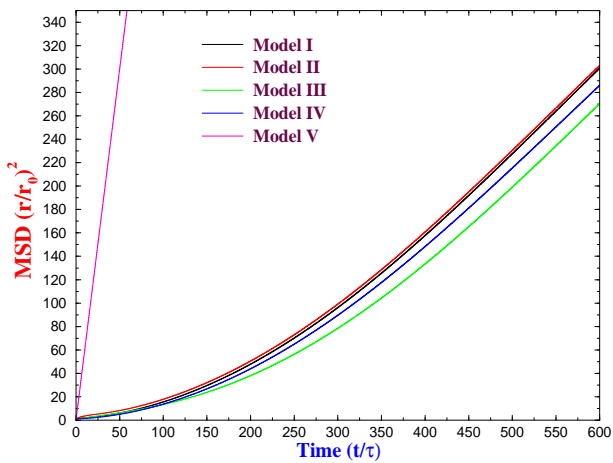,height=8cm,width=6cm,angle=-90}
\caption{Super-diffusive time dependence of $MSD$ for all the different 
models. All the models (except Model {\bf V}) show super-diffusive 
behavior. However, the overall diffusion for all the models are much less compared to 
free particle diffusion.}
\end{figure}
 The slowing down in the dynamics due to the binding with the surface is reflected in the 
self intermediate scattering function $F_{s}({\bf k},t)$. 
Figure $5$ shows the non-exponential time dependence of the $F_{s}({\bf k},t)$ for all the 
different models 
at $k = \pi/8$. For the free diffusion, the dynamics of $F_{s}(k,t)$ is exponential as 
expected. For all the other models, the non-exponentiality originates from the very slow 
relaxation of the system. The pronounced sub-diffusive behavior is reflected in 
the short time dynamics of $F_{s}(k,t)$ in case of the Model {\bf II}. Similar signature 
of sub-diffusive dynamics is observed in case the second-rank dipole-dipole correlation 
function of lysozime in water from computer simulation \cite{massimo}. Mode coupling 
theory of glassy liquids provide an explanation for the sub-diffusive behavior in terms 
of $\langle \Delta r^{2}(t)\rangle$ and self intermediate scattering function. 
{$F_{s}(k,t)$ can be expressed from Gaussian approximation as,
\begin{eqnarray}
F_{s}(k,t) = exp \biggl(-{{k^2\langle \Delta r^{2}(t)\rangle}\over{6}}\biggr)\\
= exp \biggl(-{A\, {k^2\,t^{\alpha}}}\biggr)
\end{eqnarray}
\noindent where $A$ is a numerical constant related to diffusion. 
Clearly for $\alpha$ not equal to unity, $F_{s}(k,t)$ becomes non-exponential. 
$F_(k,t)$ plotted in figure $4$ have been fitted to stretched exponential and the 
{\it exponents obtained are considerably less than unity in all the models} except for 
the free diffusion which models the bulk water is purely exponential which is consistent 
with the relaxation properties observed in bulk water \cite{bouchaud}.
\section{conclusion}
\begin{figure}[tb]
\epsfig{file=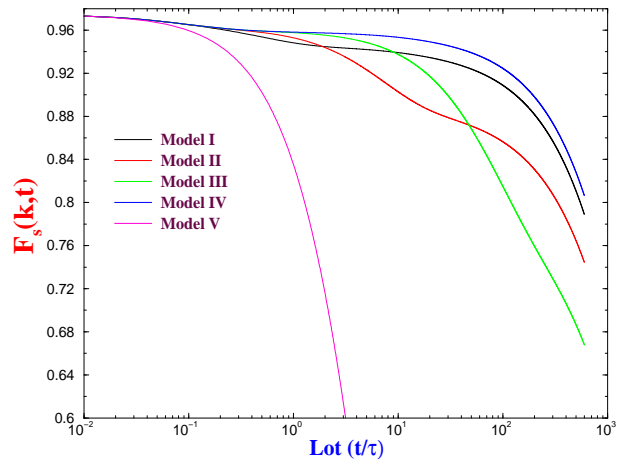,height=8cm,width=6cm,angle=-90}
\caption{Self intermediate scattering function ($F_{s}(k,t)$) is plotted against 
$Log(t/\tau)$ for all the models at $k=2\pi$. Due to slow diffusion, all the models 
(except Model {\bf V}) show non-exponential relaxation. $F_{s}(k,t)$ for Model {\bf II} 
in the short time shows the signature of sub-diffusive relaxation (see text).}
\end{figure}
 Recent observations regarding the slow orientational and translational dynamics of the 
biological water compared to the bulk water is mapped to a simple problem of diffusion 
over a double well potential. The first well corresponds to the first coordination shell 
where the water is strongly hydrogen bonded to the biological surface and enthalpically very 
stable. The second well corresponds to the relatively free water which constitute the 
second coordination shell. 
Diffusion over such a potential landscape is shown to mimic the recent experimental and 
simulation studies showing a cross-over from sub-diffusive to super-diffusive dynamics. 
{\it Sub-diffusive dynamics originates not only from the relative stability of the 
potential well, but also from the backward diffusion process from the second well to 
the first well}. The slow long time relaxation is reflected from the {\it stretched 
exponential behavior of $F_{s}({\bf k},t)$}. The present study provides convincing 
evidence that the binding to a constrained surface along with exchange between the free 
and bound water could be the primary reason for the sub-diffusive dynamics of 
biological water. In view of the results obtained in this work, the origin of 
super-diffusivity can be explained by a Heaviside step function representation of 
$MSD$ as given by, 
\begin{eqnarray}
MSD = \langle |\Delta r(t)|^{2} \rangle = \int d\epsilon P(\epsilon) 
\biggl[ H (t - k_{bf}^{-1}) \times \\ \nonumber
6 D_{T} (t -  k_{bf}^{-1} ) - a^{2}\, t \,H (k_{fb}^{-1}+ k_{bf}^{-1} - t) \biggr]
\end{eqnarray} 
\noindent where $P(\epsilon)$ is the probability distribution of the potential energy of 
binding ($\epsilon$). $a$ is the separation between two wells (around 3 \AA). 
$k_{bf}$ and $k_{fb}$ are rate of conversion from bound to free and 
free to bound, respectively. $H(t)$ is the Heaviside step function which is equal to 
unity for positive values of its argument and zero otherwise. So if the rate of 
conversion from bound to free is large, then the $MSD$ will show a super-diffusive 
behavior. However, the free to bound transition retards diffusion and gives rise to 
sub-diffusive behavior. Note that free to bound transition is faster than bound to 
free because of much lower activation barrier in the former case.

\end{document}